\documentclass[]{aa}
\usepackage{graphicx}
\usepackage{txfonts}

\newcommand{\Lsun}{\mbox{$L_{\odot}\;$}}
\newcommand{\Msun}{\mbox{$M_{\odot}\;$}}
\def\lsim{\;\raise0.3ex\hbox{$<$\kern-0.75em\raise-1.1ex\hbox{$\sim$}}\;}
\def\gsim{\;\raise0.3ex\hbox{$>$\kern-0.75em\raise-1.1ex\hbox{$\sim$}}\;}

\def\ngc{NGC~6334}

\def\etal{et al. }

\def\beq{\begin{equation}}
\def\enq{\end{equation}}
\def\begar{\begin{eqnarray}}
\def\endar{\end{eqnarray}}
\def\mathnew{\mathsurround=0pt}
\def\simov#1#2{\lower .5pt\vbox{\baselineskip0pt \lineskip-.5pt
        \ialign{$\mathnew#1\hfil##\hfil$\crcr#2\crcr\sim\crcr}}}

\def\cmc{\rm ~cm^{-3}}
\def\cms{\rm ~cm^{-2}}
\def\kms{\rm ~km~s^{-1}}
\def\ergs{\rm ~erg~s^{-1}}
\def\enf{\rm ~erg~cm^{-2}~s^{-1}}

\def\nh{N$_{\rm H}$ }
\def \rchisq {$\chi_{\nu}^{2}$}
\def \chan {{\it Chandra}\,}
\def \asca {{\it ASCA}\,}
\def \xmm {{\it XMM-Newton}\,}
\def \jemx {{\it JEM-X}\,}
\def \int {{\it INTEGRAL}\,}
\def \isgri {{\it ISGRI}\,}
\def\s1{\hspace*{-1mm}}

\def \nh {N${\rm _H}$}

\def \hcm {\hbox {\ifmmode $ atom cm$^{-2}\else atom cm$^{-2}$\fi}}
\def \arcmin {\hbox{$^\prime$}}
\def \arcsec {\hbox{$^{\prime\prime}$}}

\def \rchisq {$\chi_{\nu} ^{2}$}
\def\approxgt{\mathrel{\hbox{\rlap{\lower.55ex \hbox {$\sim$}}
        \kern-.3em \raise.4ex \hbox{$>$}}}}
\def\approxlt{\mathrel{\hbox{\rlap{\lower.55ex \hbox {$\sim$}}
        \kern-.3em \raise.4ex \hbox{$<$}}}}
\begin{document}
\title{
INTEGRAL detection of hard X-rays from NGC 6334\thanks{ The present
work is partly based on observations with INTEGRAL, an ESA project
with instruments and a science data centre funded by ESA member
states (especially the PI countries: Denmark, France, Germany,
Italy, Switzerland, Spain), Czech Republic and Poland, and with the
participation of Russia and the USA. This research has made use of
data obtained from the NASA High Energy Astrophysics Science Archive
Research Center.}:\\
Nonthermal emission from colliding winds or an AGN?}
     \author{
     A.M.Bykov
     \inst{1},
     A.M.Krassilchtchikov
     \inst{1},
     Yu.A.Uvarov
     \inst{1},
     F.Lebrun
     \inst{2},
     M.Renaud
     \inst{2},
     R.Terrier
     \inst{3,2},
     H.Bloemen
     \inst{4},
     B.McBreen
     \inst{5},
     T.J.-L.Courvoisier
     \inst{6,7},
     M.Yu.Gustov
     \inst{1},
     W.Hermsen
     \inst{4,8},
     J.-C.\,Leyder
     \inst{6,9},
     T.A.Lozinskaya
     \inst{10},
     G.Rauw
     \inst{9},
     J.-P.\,Swings
     \inst{9}
          }
   \offprints{A.M.Bykov~(byk@astro.ioffe.ru)}
   \institute{A.F.Ioffe Institute for Physics and Technology,
              26 Polytechnicheskaia, 194021, St.Petersburg, Russia 
    \and
            CEA-Saclay, DSM/DAPNIA/Service d'Astrophysique,
            91191 Gif-sur-Yvette Cedex, France 
     \and
            APC-UMR 7164, 11 Place M.Berthelot, 75231
            Paris, France 
    \and
            SRON Netherlands Institute for Space Research, Sorbonnelaan 2,
            3584 CA Utrecht, The Netherlands 
  \and
        Department of Physics, University College Dublin, Dublin 4, Ireland
 \and
        INTEGRAL Science Data Centre,
        Chemin d'\'Ecogia 16, 1290 Versoix, Switzerland
\and
         Geneva Observatory, Chemin des Maillettes 51, CH-1290 Sauverny, Switzerland
  \and
             Astronomical Institute "Anton Pannekoek", University of Amsterdam, Kruislaan 403,
             NL-1098 SJ Amsterdam, The Netherlands
\and
            Institut d'Astrophysique et de G\'eophysique, Universit\'e de Li\`ege,
            All\'ee du 6 Ao\^ut 17, B\^at B5c, 4000 Li\`ege, Belgium  
   \and
        Sternberg Astronomical Institute, Moscow State University,
        13 Universitetskij, 119899, Moscow, Russia 
            }
\date{Received September 1, 2005; accepted December 20, 2005}
\abstract
{}
{We report the detection of hard X-ray emission from the field of
the star-forming region \ngc with the the International Gamma-Ray
Astrophysics Laboratory \int.}
{The \jemx\ monitor and \isgri\ imager aboard \int\ and \chan\
{\it ACIS} imager were used to construct 3-80 keV images and
spectra of \ngc.}
{The 3-10 keV and 10-35 keV images made with \jemx\ show a complex
structure of extended emission from \ngc\s1. The \isgri\ source
detected in the energy ranges 20-40 keV, 40-80 keV, and 20-60 keV
coincides with the \ngc ridge. The 20-60 keV flux from the source
is (1.8$\pm$0.37)$\times$10$^{-11} \enf$. Spectral analysis of the
source revealed a hard power-law component with a photon index
about 1. The observed X-ray fluxes are in agreement with
extrapolations of X-ray imaging observations of \ngc by \chan {\it
ACIS} and \asca {\it GIS}.}
{The X-ray data are consistent with two very different physical
models. A probable scenario is emission from a heavily absorbed,
compact and hard \chan\ source that is associated with the AGN
candidate radio source NGC 6334B. Another possible model is the
extended \chan\ source of nonthermal emission from \ngc\ that
can also account for the hard X-ray emission observed by {\it
INTEGRAL}. The origin of the emission in this scenario is due to
electron acceleration in energetic outflows from massive early
type stars. The possibility of emission from a young supernova
remnant, as suggested by earlier infrared observations of \ngc\s1,
is constrained by the non-detection of $^{44}$Ti lines.}
\keywords{Gamma rays: observations --- X rays: ISM --- Star
Forming Regions --- individual: \ngc }
\authorrunning{A.M.Bykov et al.}
\titlerunning{Hard X-ray Emission from \ngc}
\maketitle
\section{Introduction}
\ngc is a star forming (SF) complex of a total bolometric luminosity
$\sim$1.9$\times$10$^6 \Lsun$ associated with a giant molecular
cloud of mass $\sim$1.6$\times$10$^5 \Msun$ (e.g. Loughran \etal
1986). The complex structure of \ngc with a number of clearly
separated and localized SF sites was established by radio and
infrared (IR) observations. Massive SF sites reside along a ridge of
size about 20$\arcmin \times $3$\arcmin$ ($\sim$ 10 $\times$
1.5 pc) and are associated with the peaks in the main ridge of
molecular gas. The ridge can be seen in many energy bands ranging
from 843 MHz radio to several keV X-rays.
The standard tracers of SF regions such as [C~II] 158$\mu$m,
[O~I] 145$\mu$m, and  [O~I] 63$\mu$m, as well as rich molecular
emission spectra revealed structured emission of \ngc with
clumps of far-infrared (FIR) sources. Recently Kraemer and Jackson
(1999) reviewed molecular gas in \ngc and constructed detailed
maps of the CO, CS, and NH$_3$ emission regions. They found a complex
structure of the gas distribution with a number of molecular
filaments and bubbles. High resolution VLA observations of \ngc by
Carral \etal (2002) revealed shell-like structures of diameters from
0.12 to 3.5 pc probably tracing stellar winds. Nonthermal radio
emission features, H$_2$O, OH, and methanol maser sources, as well as high
magnetic fields (about 200 $\mu$G) were observed in \ngc (e.g. Sarma
\etal 2000 and references therein). We adopt the distance of
$\sim$~1.7~kpc to \ngc (Neckel 1978).

McBreen \etal (1979)
discovered a bright ($\sim$1.9$\times$10$^5$ \Lsun bolometric
luminosity) source of IR emission in the southwestern part of the SF
ridge which was not detected in previous IR observations of the
region. The source position is associated with H$_2$O and OH masers.
Bipolar structure of extended H$_2$ emission and an energetic
outflow were found in the source (as well as in a nearby FIR
source). The nature of the apparent variability was attributed by
McBreen \etal (1979) to a sudden luminosity onset during early SF
phenomena or, alternatively, to a supernova (SN) hidden in a
molecular cloud. Young SN remnants (SNRs) possibly hidden in dense
molecular clouds can be identified by their gamma-ray emission
lines from $^{44}$Ti with the \isgri\ detector (Lebrun \etal 2003).
Hard X-ray observations are a perfect tool to look for a SNR hidden
in a dense molecular cloud. Being compact sites of powerful
kinetic energy release and fast shock waves generated by stellar
winds and SNRs, massive SF regions are expected to have non-thermal emission
components (e.g. Bykov 2001). This possibility strongly
motivates \int\ data analysis of \ngc\s1.

\ngc was observed with \asca\ (Matsuzaki \etal 1999;
Sekimoto \etal 2000). X-ray emission was detected from the SF ridge
and from the source AXJ~1720.3-3544 that is about 10\arcmin\ to the
north of the center of the ridge. The emission spectrum of the
ridge sources was fitted with a thermal hot plasma of temperature
$\sim$ 9 keV. The source AXJ~1720.3-3544 was identified as the
B0.5e star CD-35 11482. Recently Ezoe \etal (2005) analyzed \chan\
data on \ngc\s1. In addition to 800 point sources in the field they
found a diffuse X-ray emission region of 5$\times$9 pc size and of luminosity
2$\times 10^{33} \ergs$. The authors suggested that thermal
emission of several keV is due to plasma heated by stellar wind shocks,
while a flat continuum is due to accelerated particles.

However, it is not easy to distinguish hot thermal gas from
nonthermal emission generated by accelerated particles with an
energy band that is limited to 10 keV. We present below the first
hard X-ray (3-80 keV) images and spectra of \ngc made with \int.

\begin{figure*}
\includegraphics[width=1.0\textwidth]{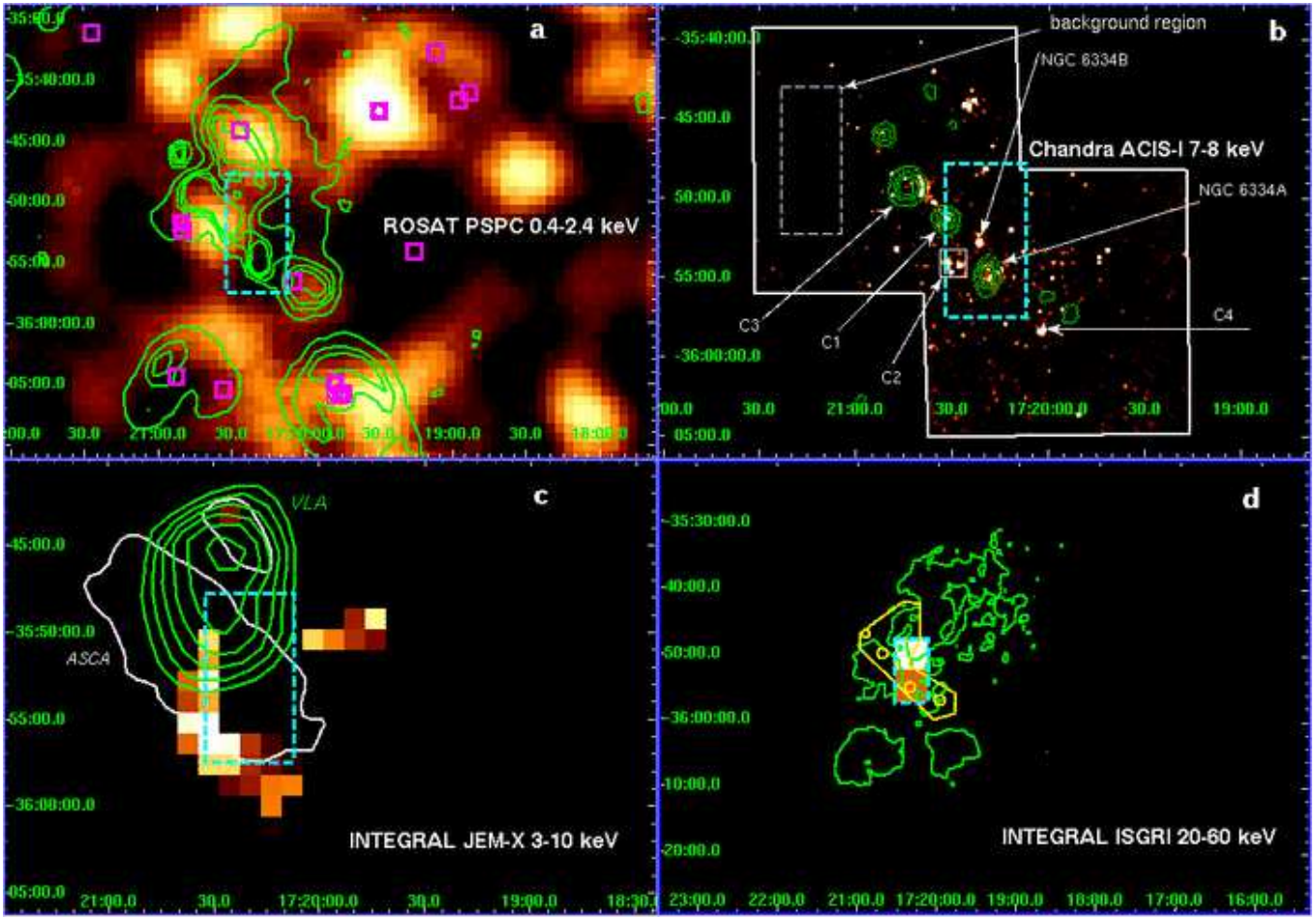}
\caption{a) {\it ROSAT} RASS 0.4-2.4 keV map with {\it
MOST} MGPS 843 MHz contours. Massive stars, visible in the optical
band are shown by boxcircles (O stars) and boxes (B stars). 
b) \chan\ 7-8 keV map smoothed with a 2 pixel Gaussian kernel
and {\it VLA} NVSS 1.4 GHz contours.
The background region used for \chan\ analysis is marked as well as the main
sources of \chan\ emission within the \isgri\ excess.
c)~\jemx\ 3-10 keV map with \asca {\it GIS} 6-10 keV contours
and {\it VLA} GPS 8.35 GHz contours. 
d) \isgri\ 20-60 keV map with DSS-R contours. The P-shaped
region with circles indicates the \ngc\ ridge with SF complexes as
seen in 71$\mu$m IR band by Loughran et al. (1986). Bright \isgri\
pixels are shown on panels a) -- c) as dashed rectangles.}
\label{fig:image}
\end{figure*}

\section{Observations and Data Analysis}

\subsection{\int\ observations}

The regions surrounding \ngc have been observed with the
\isgri\ imager (Ubertini \etal 2003, Lebrun \etal 2003) and \jemx\
monitor (Lund \etal 2003) aboard \int\ (Winkler \etal 2003).

We analysed 420 ks of fully coded field of view (FCFOV) data from
\isgri and 92 ks of FCFOV data from \jemx. The data were taken
during revolutions 46 -- 167 (28 Feb 2003 -- 27 Feb 2004) and are
now public.

The \int\ data have been reduced with the standard off-line scientific analysis
 OSA 5.0 version (Courvoisier \etal 2003).
The standard good time selection criteria were applied; only science
windows with more than 100~s of good time were considered.

The \int\ images of the field around \ngc\ reveal the complex
morphology of the X-ray emission. Archival {\it ROSAT}, \asca,
\chan\ and {\it CGRO EGRET} observations of \ngc were analyzed in
addition to the \int\ data. Available archive {\it MOST}, {\it VLA},
{\it IRAS}, {\it MSX}, 2MASS, and DSS data was used to construct a
multiwavelength picture of a few degrees  in size around the
\ngc\ complex.

Hard X-ray emission from \ngc can be efficiently studied by
combining \isgri\ and high resolution \chan\ observations. We used
 two \chan\ 40 ksec observations (ObsIDs 2573 and 2574) of
\ngc\ performed on 31 Aug and 2 Sep 2002 (P.I. Y.Ezoe). The
source was imaged with four ACIS-I CCDs. The event files were
processed with the standard {\it CIAO} tools (ver.\ 3.1, CALDB 2.28).

In Fig.~\ref{fig:image} we show multiwavelength images of NGC 6334.
The \ngc ridge, as seen in IR, is shown on panel
{\bf d}, while its counterparts at 843 MHz and 1.4 GHz radio
and in keV X-rays are visible on panels {\bf a, b} and {\bf c}. The
morphology of the 10-35 keV \jemx\ image is very similar to that of
3-10 keV; the 20-40 keV and 40-80 keV maps by \isgri\ are also very
similar to the 20-60 keV map shown in Fig.~\ref{fig:image}d. The
apparent \jemx\ NGC 6334B counterpart is $\sim$ 2.5\arcmin\ from the Chandra
position (within the \jemx\ resolution of 3\arcmin).

The {\it ROSAT} RASS map shows a patchy structure caused by
the distribution of X-ray sources and strong local \nh variations
in the molecular cloud complex. The 7-8 keV map of \chan\ consists of
extended emission regions correlated with the 1.4 GHz radio
excesses.

From cross calibration with the Crab we derived the following flux
estimates for the \isgri\ source: (1.8$\pm$0.37)$\times$10$^{-11}
\enf$ in the 20-60 keV band, (1.1$\pm$0.26)$\times$10$^{-11} \enf$
in the 20-40 keV band, and (1.2$\pm$0.39)$\times$10$^{-11} \enf$
in the 40-80 keV band. The significances of the detections are
4.9, 4.2, 3.1 $\sigma$, respectively.

\subsection{Spectral analysis of the hard X-ray emission}

The modest resolution of the \int\ high energy images (panels {\bf
c} and {\bf d} in Fig.~\ref{fig:image}) strongly motivates a joint
analysis of high resolution \chan\ observations with that of \int\, to
constrain the nature of the nonthermal emission. We made a spectral
analysis of the \chan\, 0.8-8.0 keV emission extracted from the
rectangular \isgri\ excess region of size 5\arcmin $\times$
10\arcmin, shown in Fig.~\ref{fig:image}{\bf a, b} and {\bf c}.

Inspection of the \chan\ sources in the field revealed two
probable scenarios. We find that the observed \isgri\ flux may
come from a bright point-like hard \chan\ source located at (FK5:
17:20:21.8; -35:52:48) that is clearly seen in
Fig.\ref{fig:image}b. The source coincides with the known radio
source NGC 6334B  (FK5: 17:20:21.8; -35:52:48.2) studied by Moran
and collaborators (see Moran \etal (1990) and references therein).
They suggested that NGC 6334B is a background extragalactic
source.

Alternatively, the \isgri\, hard emission may originate from the
extended rectangular region covered by the two bright \isgri\
pixels, with the emission within a radius of 10\arcsec from NGC
6334B subtracted. We will now refer to it as the hard extended
emission (HEE) region. The HEE region is not dominated by one or
two bright point-like \chan\ sources, but it has a structured
morphology with a few bright clumps. The clumps may be either
extended or consist of many weak sources. In the HEE region we
further selected three extended regions associated with the X-ray
emission clumps. They are marked as NGC 6334A, C1, and C2 in
Fig.~\ref{fig:image}{\bf b}. The sources NGC 6334A and C1 are
bounded by an NVSS radio contours shown in Fig.~\ref{fig:image}{\bf
b}. NGC 6334A contains a radio shell of 15\arcsec diameter and of
about 10 Jy flux density at 6 cm found by Carral \etal (2002). The
HII region NGC 6334A is coincident with a bright IR source
(McBreen \etal 1979). In Fig.~\ref{fig:NGC6334A} we present a
zoomed \chan image of the region. We found no clear \chan
counterpart to the central source of the NGC 6334A radio shell.
The central source of 6.1$\pm$0.9 mJy flux density at 3.5 cm was
found by Carral \etal (2002) in high resolution VLA data. However,
there is a possible \chan counterpart to the IRS 19 source also
present within the radio shell (see Fig.~\ref{fig:NGC6334A}).

\begin{table}
\caption{Parameters of 0.8-8.0 keV \chan fits for sources within the HEE
region shown in Fig.\ref{fig:image}b.}
\label{tab:chandra}
\begin{tabular}{c|cccc} \hline
Source       & NGC 6334B   & NGC 6334A           &  C1           &  C2            \\ \hline
{N$_H$}$^a$  & 13 $\pm$ 3.0 & 2.7 $\pm$ 0.5 & 0.6 $\pm$ 0.1 &  0.5 $\pm$ 0.1  \\
$\Gamma$     & 0.9 $\pm$ 0.5 & 1.4 $\pm$ 0.3 & 1.7 $\pm$ 0.1 & 1.5 $\pm$ 0.1  \\
$\chi_{\nu}^2$/dof & 19/47  & 30/51         & 99/92          & 48/62          \\
{F$_x$}$^{b}$ & 2.9         & 1.0           & 0.9            & 0.6            \\ \hline
\end{tabular}
\begin{tabular}{l}
$^a$ Hydrogen column density, in 10$^{22}\cms$ \\
$^b$ Unabsorbed flux in 2.0--7.0 keV band, in $10^{-12}$ erg cm$^{-2}$ s$^{-1}$
\end{tabular}
\end{table}

\begin{figure*}
\includegraphics[angle=270,width=0.5\textwidth,bb=105 45 558 709,clip]{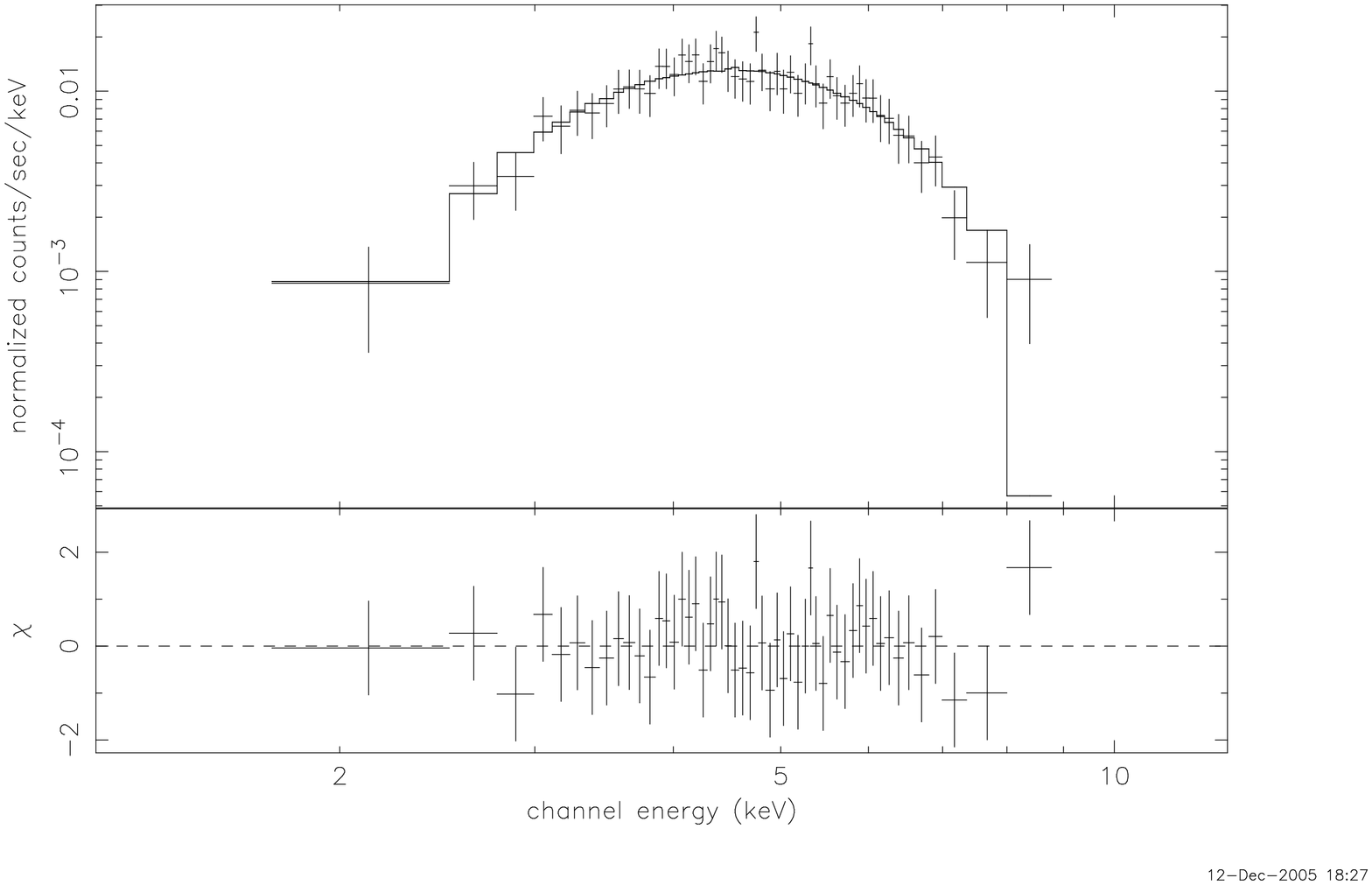}
\includegraphics[angle=270,width=0.5\textwidth,bb=105 45 558 709,clip]{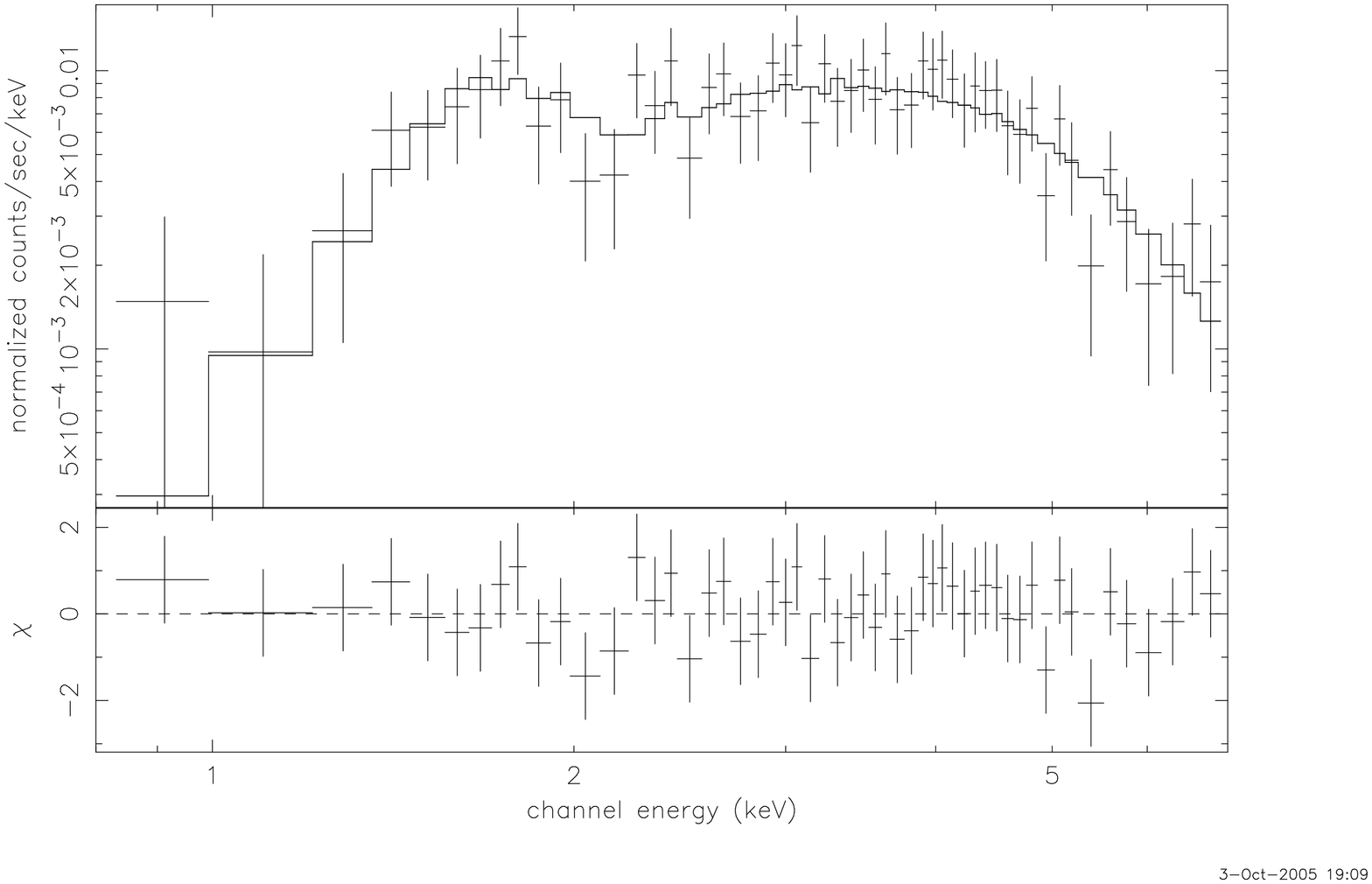}
\caption{\chan\ spectra of NGC 6334B (left panel) and of
NGC 6334A (right panel). Fits are shown as solid lines and have
\rchisq $\lsim$ 1.} \label{fig:chanSpec}
\end{figure*}

\begin{figure*}
\includegraphics[angle=270,width=0.5\textwidth,bb=95 47 560 707,clip]{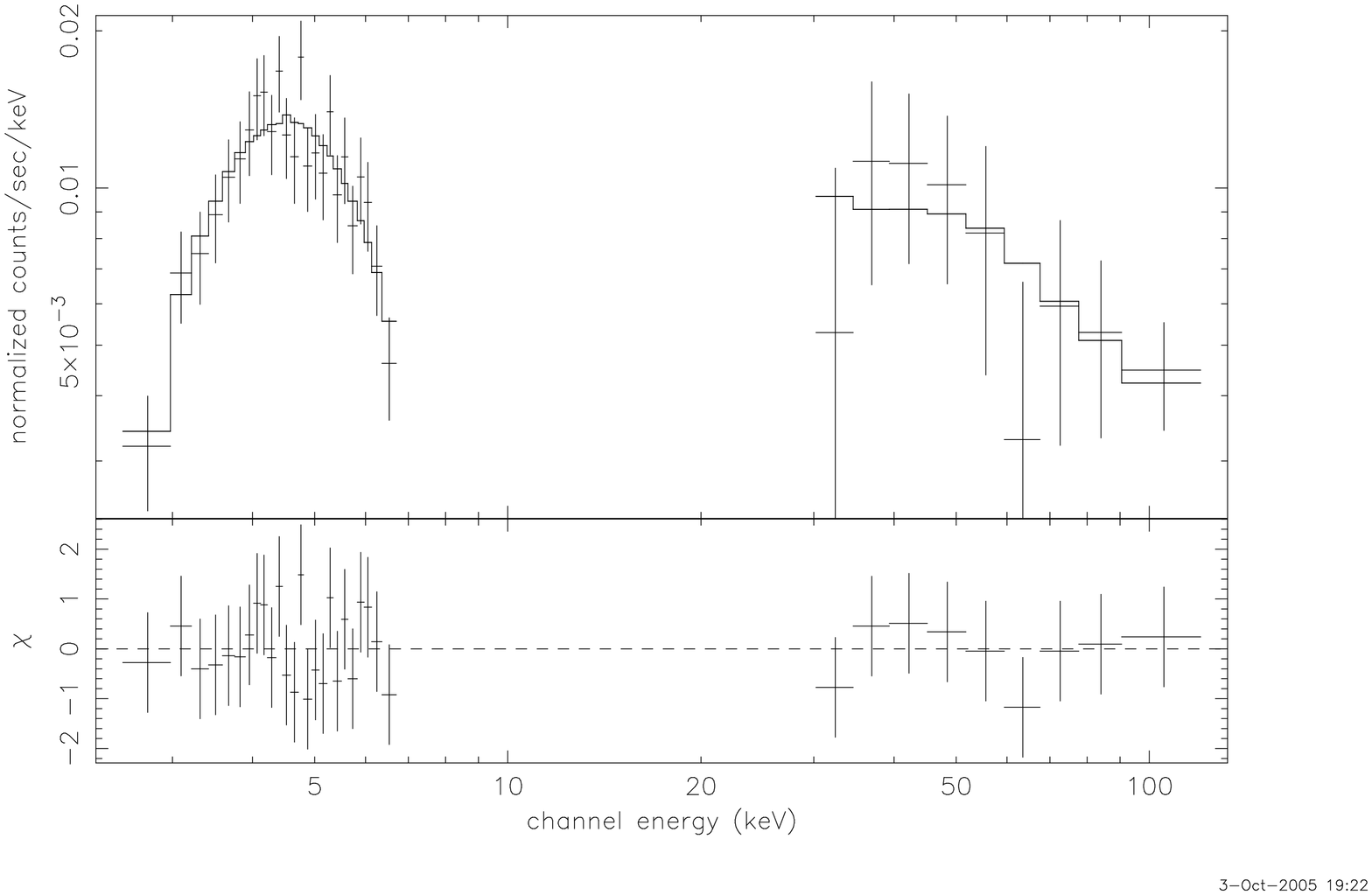}
\includegraphics[angle=270,width=0.5\textwidth,bb=95 47 560 707,clip]{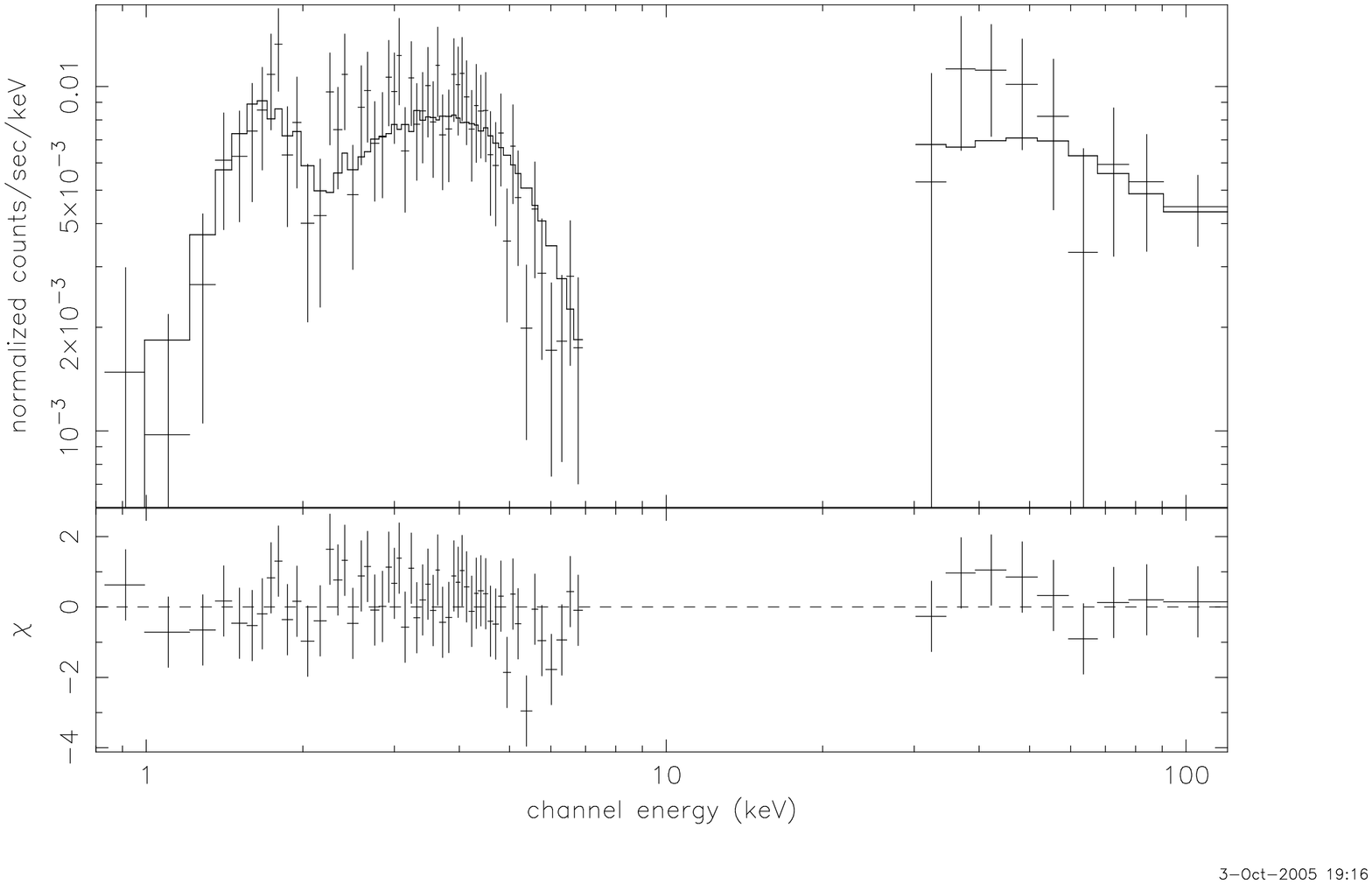}
\caption{The combined \chan-\isgri spectra of NGC 6334B (left panel) and
of NGC 6334A (right panel).
Fits are shown as solid lines and have \rchisq $\lsim$ 0.75.}
\label{fig:spectra}
\end{figure*}

For \chan\ spectral analysis we used only the northern \chan\
pointing (ObsID 2574), because the other one is contaminated by
flares. The background region used is shown in
Fig.~\ref{fig:image}b. We also used blank-sky background models
provided by the {\it CIAO}. The results of \chan\ spectral analysis
of the selected sources within the HEE region are summarized in
Table~\ref{tab:chandra}. The \chan\ spectra of the most
prominent regions NGC 6334B and NGC 6334A are shown in
Fig.~\ref{fig:chanSpec}. The combined \chan\--\isgri\ spectra
of these two sources are shown in Fig.~\ref{fig:spectra}.

The combined \chan\--\isgri\ data for NGC 6334B are fitted with
\rchisq=0.52 by a
power-law with \nh$ = (1.2 \pm 0.1)\times 10^{23}\cms$,
photon index $\Gamma$ = 0.9$\pm$0.1 and
normalization K = (2.9$\pm$0.9)$\times$10$^{-4}$ ph cm$^{-2}$
s$^{-1}$ keV$^{-1}$ at 1 keV.
The combined \chan\--\isgri\ data for NGC 6334A are fitted with
\rchisq=0.75 by a
power-law with \nh$ = (1.5 \pm 0.2)\times 10^{22}\cms$,
photon index $\Gamma$ = 0.5$\pm$0.1 and
normalization K = (6.1$\pm$1.4)$\times$10$^{-5}$ ph cm$^{-2}$
s$^{-1}$ keV$^{-1}$ at 1 keV.

The \chan\ spectrum of the whole HEE region is shown in
Fig.~\ref{fig:spectraHEE}. The spectrum contains an iron line
feature at (6.5 $\pm$ 0.1) keV. The \chan\ data are consistent with
the \asca\ spectrum of the combined FIR cores I-V reported by
Sekimoto \etal (2002) to be fitted by  a Raymond-Smith temperature
kT = 9.0$^{+3.3}_{-1.8}$ keV and a line at (6.6 $\pm$ 0.1) keV.
Alternatively, the \chan\ spectrum of the HEE region can be
modeled by a power-law with the photon index $\Gamma$ =
1.5$\pm$0.1, and N$_H$ = (1.3$\pm$0.2)$\times$10$^{22} \cms$.
In the latter case, the iron line has a non-thermal origin.
The 2-7 keV unabsorbed flux of the HEE region is 4.4$\times$10$^{-12}
\enf$. Analysis of the \chan\ spectra of the extended HEE region 
above 7 keV depends strongly on the applied background model due to
substantial spatial variations of \chan\ background above 7 keV. 
The background obtained from regions of different sizes and shapes 
was applied to the spectrum of the HEE region and yielded 
different values for the countrate in the 7-10 keV energy bin. 
The 7-10 keV rate plotted in Fig.\ref{fig:spectraHEE} was obtained 
for the blank sky{\footnote {ftp://cda.harvard.edu/pub/arcftp/caldb/}}
background and thus should be regarded as an upper limit. 
The limit is compatible with the hypothesis that the
origin of a substantial part of the 20-60 keV emission from \ngc\ 
is from the HEE region. The \chan\ spectrum of the point-like source 
NGC 6334B extends up to 9 keV (see Fig.\ref{fig:chanSpec} left panel) 
and is also well matched to the observed \isgri\ flux.  
Thus, more observations (e.g. with {\it XMM-Newton})
are required to get a real constraint on the relative contribution of 
the HEE region and the AGN-like NGC 6334B source to the \isgri\ excess.  

The five FIR core sources along the ridge of emission in \ngc\ 
were detected in the \chan\ images and have hard emission
components above 5 keV. 
To estimate their relative contribution to the hard X-ray emission 
of \ngc\ we compare their fluxes in the 5-7 keV band where 
the detection is still significant. More than a half of the hard 
X-ray emission of the whole NGC 6334 ridge (unabsorbed flux  
$\sim 2 \times 10^{-12}\enf$ in the 5-7 keV regime) comes from 
the point-like source NGC 6334B and the HEE
region. FIR cores II and V (indicated as C3 and C4 in
Fig.1b) are situated just a few arcminutes outside the HEE region
and contribute about ($\sim 0.5\times 10^{-12}\enf$) and ($\sim
0.3\times 10^{-12}\enf$), respectively, to the 5-7 keV emission of
the whole ridge. Given the proximity of the sources C3 and C4 to
the HEE region it is not possible with the angular resolution of the 
\isgri\ camera to make a meaningful conclusion on the actual contribution
of the FIR cores II and V to the observed hard emission above 20
keV. 

We have analysed the archival {\it CGRO-EGRET} data and derived
the following upper limits for gamma-ray emission from \ngc: F $<$
9.9$\times$10$^{-8}$ ph cm$^{-2}$ s$^{-1}$ for E $>$ 100 MeV, and
F $<$ 1.2$\times$10$^{-8}$ ph cm$^{-2}$ s$^{-1}$ for E $>$ 1000
MeV. These limits imply a break in the spectrum of \ngc at some
energy above 100 keV.

\section{Discussion}

From the spectral analysis of {\it IBIS-ISGRI} data we can
conclude that the hard X-ray emission has a nonthermal nature. The
X-ray data are consistent with two different physical models, or
with their combination. A plausible scenario is emission from
a heavily absorbed, compact and hard \chan\ source that is
probably associated with the obscured accreting source NGC6334B.
An alternative model is emission from the hard extended \chan\ sources
related to NGC 6334A and some other extended structures in the NGC~6334 SFR.
The origin of the emission in this scenario is due to electron
acceleration in energetic outflows from massive early type stars.

\subsection{An obscured AGN source}

A possible interpretation of the observed hard X-rays could be an
obscured AGN. The \chan\ source position coincides with better
than 1\arcsec\ accuracy with the NGC 6334B radio source studied by
Moran \etal (1990). The HI observations of NGC 6334B revealed a
prominent feature with large negative velocity at about 100 $\kms$
indicating that the distance to the source must be at least 6 kpc.
Moran \etal (1990) suggested that NGC 6334B is unrelated to the
NGC 6334 molecular cloud complex and is likely to be extragalactic.
They also summarized measurements of the $\lambda$ = 6 cm flux variability
with the flux variation from 0.18 Jy on Jun 19 1978 to 0.79 Jy on
Feb 20 1981 with the measured fluxes above 0.4 Jy  till Jul 1988.
We searched for \isgri\, flux variations in the bright 5\arcmin
pixel at the NGC 6334B position. Given the limited count
statistics of the \isgri\, source we grouped events within three
time intervals to make the light curve presented in
Fig.\ref{fig:varI}. No significant flux variations in 20-40 keV
band on a year timescale were found. The data allows to exclude
flux variations above a factor $>$ 3 on the year time scale. We
also made a light curve for \chan\ 0.8-8.0 keV flux for NGC 6334B
presented in Fig.\ref{fig:varC}. No significant flux
variations on the hour time scale were found.

Simultaneous radio and X-ray  observations can be used to
distinguish between two different types of obscured galactic
sources (see e.g. Fender 2005 for a review). Radio fluxes of
galactic black hole binaries in the low-hard states with an X-ray
flux $\lsim$ 10 mCrab (2-10 keV) are below 10 mJy. The radio
fluxes of $\gsim$ 100 mJy level were detected only in the
high-soft states with 2-10 keV X-ray flux $\sim$ 1 Crab. The radio
fluxes of some observed X-ray binaries with neutron stars are only
below that with black holes (e.g. Muno \etal 2005). We have no
simultaneous X-ray and VLA radio observations of NGC 6334B. The
available radio fluxes $\sim$ 0.4 Jy (at 6 cm) measured two
decades before the \chan\ observation are much higher than that
expected for a galactic black hole binary in a low-hard state with
$\sim$ mCrab flux in 2-10 keV band. The radio flux can be achieved
only in a relatively rare high-soft state of a stellar mass black
hole binary. Simultaneous X-ray and radio observations of
NGC~6334B are needed to finally establish its nature.

The broadband spectral characteristics of NGC 6334B given above are
consistent with that observed in obscured accreting sources
associated with AGNs (see e.g. Fabian 2004).
The absorbing column density of NGC 6334B \nh$ \sim
10^{23}\cms$ is substantially higher than typical values for the
\ngc field, where \nh$ \lsim$ 3$\times$10$^{22}\cms$.

Redshifted Fe line complexes are often observed in accreting
AGN sources. A dedicated \xmm\ observation of NGC 6334B
can be used to search for the spectral feature and thus to
constrain the source redshift. An \xmm\ observation would also
yield a significantly improved X-ray spectra for NGC 6334A
and the HEE region above 7 keV.

\subsection{Nonthermal emission of the SFR}

SF complexes are expected to have a number of high energy emission
sources. Young massive stars with stellar winds, SNRs interacting
with molecular clouds, young stellar objects with magnetic
activity, jets and Herbig-Haro outflows are among the most
plausible candidates. A number of bubbles and shells of different
sizes are apparent in the multiwavelength images in
Fig.~\ref{fig:image} and some of these are identified with winds
from young massive stars (e.g. Carral \etal 2002). Moreover, the
variability of a powerful FIR source (McBreen \etal 1979) in a close
proximity to the \isgri\ excess makes it reasonable to consider a
buried SNR as a plausible scenario for the HEE. All these objects
have to be considered as candidates for the observed hard
nonthermal emission.

The observed \isgri\ spectrum below 100 keV is well fitted by a hard
power law of photon index $\approx$ 1. We also obtained a
3$\sigma$ upper limit for the 67.9 keV and 78.4 keV $^{44}$Ti
lines of 2.1$\times$10$^{-5}$ ph cm$^{-2}$ s$^{-1}$. This implies
a maximal $^{44}$Ti yield of 2.3$\times$10$^{-6} \Msun$ which is much
lower than predicted by most of the current core-collapse SN
models for a SN of age $<$ 100 years (e.g. Woosley and Weaver
1995).

\begin{figure}
\includegraphics[angle=0,width=0.5\textwidth]{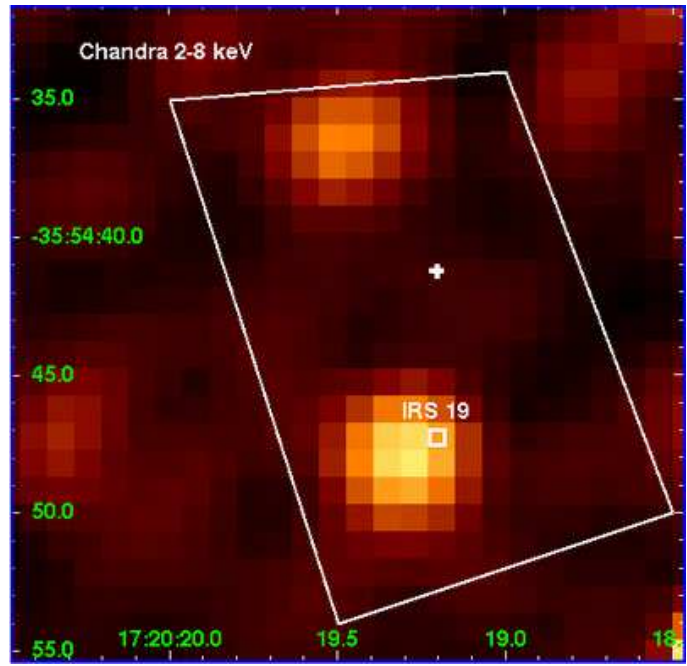}
\caption{\chan\ 2.0-8.0 keV image of NGC 6334A. Position of
the central point VLA source (taken from Carral \etal (2002))  is
marked by a cross, while that for the IRS 19 is marked by a box,
the boundary of the VLA radio shell is indicated as a
parallelogram-like frame.} \label{fig:NGC6334A}
\end{figure}

\begin{figure}
\includegraphics[angle=270,width=0.5\textwidth,bb=20 75 595 800,clip]{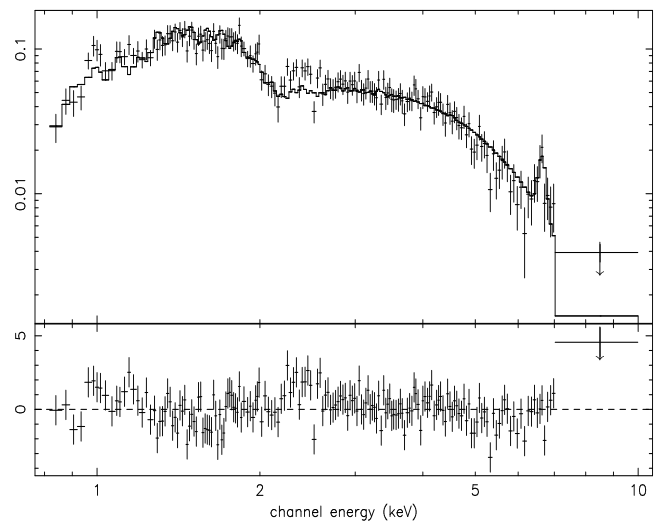}
\caption{\chan spectrum of the HEE region.}
\label{fig:spectraHEE}
\end{figure}

\begin{figure}
\includegraphics[angle=270,width=0.5\textwidth]{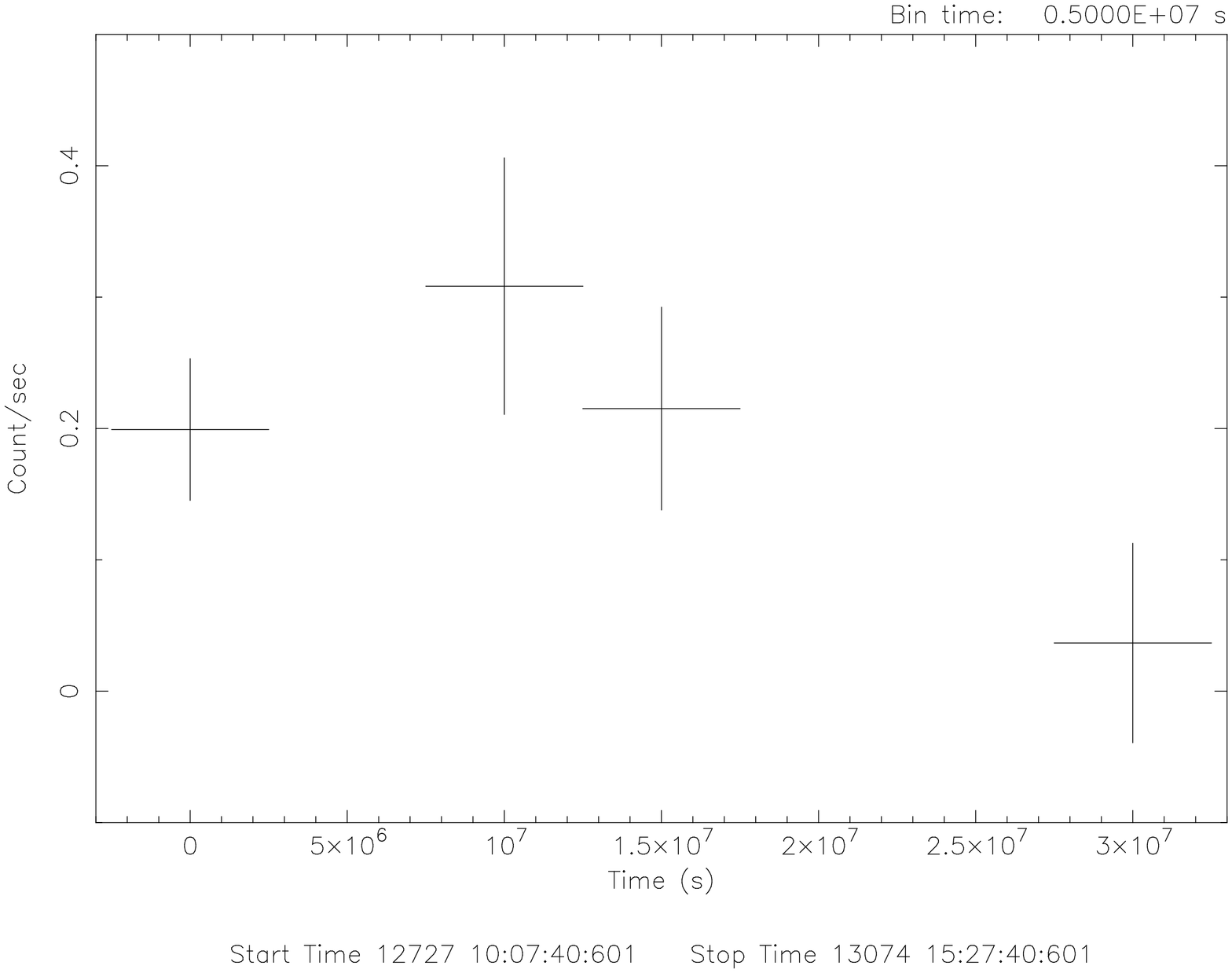}
\caption{The 20-40 keV \isgri\ lightcurve of NGC 6334B. The
bounding dates are in the Truncated Julian Days (TJD) defined as
TJD~=~JD~-~2~440~000.5.} \label{fig:varI}
\end{figure}

\begin{figure}
\includegraphics[angle=270,width=0.5\textwidth]{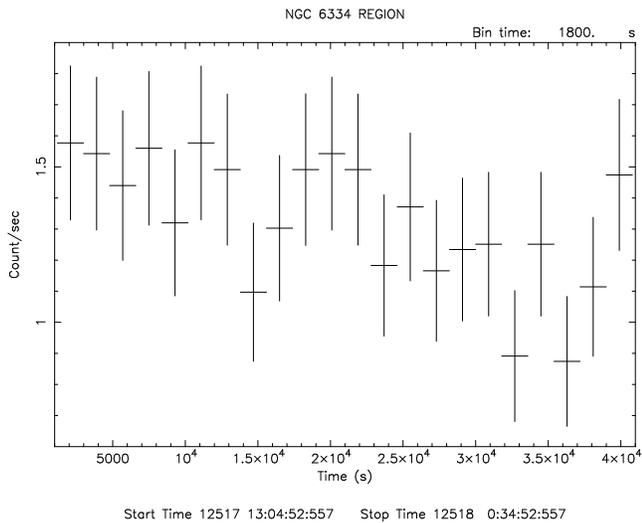}
\caption{The 0.8-8.0 keV \chan\ light curve of NGC 6334B. The
bounding dates are in the Truncated Julian Days (TJD) defined as
TJD=JD-2~440~000.5.} \label{fig:varC}
\end{figure}

The powerful (of L$_b \sim $2.5$\times $10$^5~\Lsun$ bolometric
luminosity) FIR bipolar source with a high velocity and a high
luminosity H$_2$O maser, some Herbig-Haro objects (e.g. Loughran
\etal 1986), and a well defined $\sim 15\arcsec$ shell that may be
an O7.5 star are located in the region (Carral \etal 2002). The
VLA radio shell is indicated as a parallelogram frame in the \chan
image in Fig.~\ref{fig:NGC6334A}. The mechanical luminosity of the
O7.5 star wind was estimated by Carral \etal (2002) to be $L_w
\sim $10$^{37}~\ergs$ i.e. a few percent of L$_b$. If the
bolometric luminosity of the source was initially released as a
mechanical luminosity of supersonic outflows from young
(proto)stellar source population and then converted into the FIR
emission then the hard X-ray emission can be produced by
bremsstrahlung emission of fast electrons (of energies below MeV
regime) accelerated in the outflows. The efficiency of X-ray
bremsstrahlung emission is relatively low due to the Coulomb
losses of fast particles, providing powerful IR extended emission
from the dense medium. The conversion efficiency of the
bremsstrahlung emission is $\eta \sim $10$^{-5}$ in an ambient
medium [of solar abundance] for the energy band below MeV. One may
conclude that hard X-ray emission of L$_x \lsim $10$^{34} \ergs$
can still be produced by the MeV electrons. A model of electron
acceleration by MHD shock waves with velocities about 100 $\kms$,
that are typical for sub-parsec scale size outflows  in a dense
ambient medium with n$_a \gsim $10$^4 \cmc$, was made by Bykov
\etal (2000). They considered the case of a supersonic flow
produced by a supernova shock in a dense clump of a molecular
cloud, but the model can be applied to a MHD shock of a similar
size in a dense medium produced by any other supersonic outflow,
e.g. by a fast wind in the dense medium. A distinctive feature of
the model is a very flat X-ray spectrum of photon index $\Gamma
\sim $1. An energetically more efficient model of the hard X-ray
emission with the conversion efficiency $\eta \gsim $10$^{-3}$
could be realized in case of synchrotron emission of TeV regime
electrons accelerated by fast ($\gsim 1,000 \kms$) shock waves 
in colliding winds.

Given the high significance of the \chan\ detection of both AGN
and HEE sources it is also possible that both are contributing to
the observed hard X-ray emission. Future high resolution
observations with
{\it Simbol-X} (Ferrando \etal 2004) or {\it Gamma Ray Imager} could 
help to resolve the issue.

\section{Conclusions}

1. We detected a hard X-ray emission source in the galactic
star-forming region \ngc with the \jemx and {\it IBIS/}\isgri\
telescopes aboard the International Gamma-Ray Astrophysics
Laboratory \int. The source has a nonthermal spectrum at least up
to 100 keV.

2. From the multiwavelength analysis of the complex \ngc region we
concluded that the source may be associated both with the background,
likely extragalactic, radio source NGC 6334B projected onto the
\ngc SFR, and with an extended HII region, associated with bright IR
source and a radio shell NGC 6334A.

\begin{acknowledgements}

We thank the anonymous referee for constructive comments.
The work was partially supported by RFBR grants 03-02-17433,
04-02-16595, 04-02-16042, 03-07-90200, RAS program,
Russian Leading Scientific Schools grant 1115.2003.2,
and by the ESA.
JCL, GR and JPS acknowledge support through the XMM-INTEGRAL
PRODEX project, the Belgian FNRS, and IAP contract P5/36. Support
from the International Space Science Institute (Bern) through the
international teams program is gratefully acknowledged.

\end{acknowledgements}

{\bf References} \\

Bykov, A.M., Chevalier, R.A., Ellison, D.C. \& Uvarov, Yu.A. 2000,
ApJ, 538, 203

Bykov, A.M. 2001, Space Sci. Rev., 99, 317

Carral, P., Kurtz, S.E., Rodriguez, L.F., et al. 2002, AJ, 123, 2574

Courvoisier, T.J.-L., Walter, R., Beckmann, V. et al. 2003, A\&A, 411, L53

Ezoe, Y., Kokubun, M., Makishima, K., 2005 in: Star Formation in
the Era of Three Great Observatories, p.49

Fabian, A.C.  2004,  in: Coevolution of Black Holes and Galaxies.
Ed. L.Ho,  Cambridge University Press, 447

Fender, R.P.  2005,  in: Compact Stellar X-ray Sources. Ed. W.H.G.
Lewin and M. van der Klis, Cambridge University Press, 2005,
[astro-ph/0303339]

Ferrando, Ph., Arnaud, M., Cordier, B. et al. 2004, Proc. SPIE, 5168, 65

Kraemer, K.E. \& Jackson, J.M., 1999,  ApJS, 124, 439

Lebrun, F., Leray, J.P., Lavocat, P.,  et al., 2003, A\&A, 411,
L141

Loughran, L., McBreen, B., Fazio, G.G., et al., 1986, ApJ, 303, 629

Lund, N., Budtz-Joergensen, C., Westergaard, N.J. et al. 2003, A\&A, 411, 231L

Matsuzaki, K., Sekimoto, Y., Kamae, T. et al. 1999, Astronomische Nachrichten,
320, no. 4, 323

McBreen, B., Fazio, G.G., Stier, M., Wright E.L. 1979, ApJ, 232, L183

Moran, J.M., Rodriguez, L.F., Greene, B. et al. 1990, ApJ, 348, 147

Muno, M.P., Belloni, T., Dhawan, V.,  \etal 2005, ApJ, 626, 1020

Neckel, T., 1978,  A\&A, 69, 51

Sarma, A.P., Troland, P.H., Roberts, D.A. et al. 2000, ApJ, 533, 271

Sekimoto, Y., Matsuzaki, K., Kamae, T. et al. 2000, PASJ, 52, L31

Ubertini, P., Lebrun, F., Di Cocco, G. et al. 2003, A\&A, 411, L131

Winkler, C., Courvoisier, T., Di Cocco, G.,  et al. 2003, A\&A,
411, L1

Woosley, S.E., Weaver, T.A. 1995, ApJS, 101, 181

\newpage

\begin{figure}
\includegraphics[width=\textwidth]{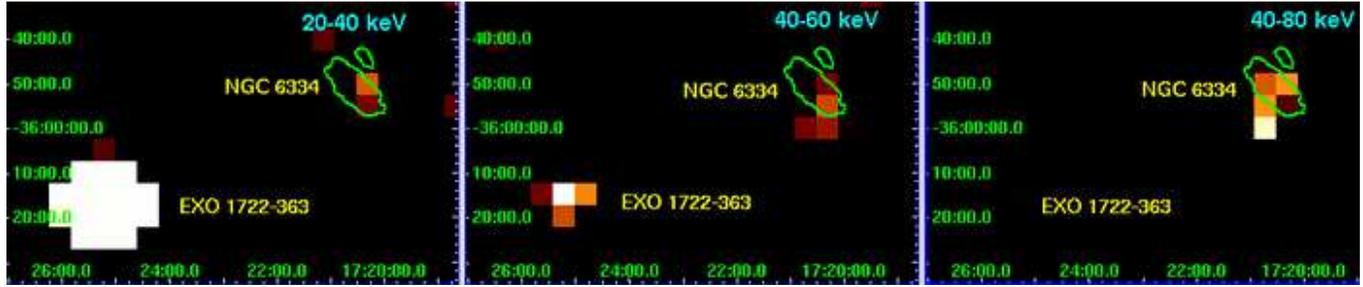}
\caption{ A broad multiband {\it ISGRI}  view of the field of NGC 6334. The 
neighbouring source is EXO 1722-363.
The hard source in NGC 6334 keeps visible up to the 40-80 keV band, while 
the bright HMXB (see Corbet, R.H.D., Markwardt, C.B., and Swank, J.H., 
2005, ApJ, {\bf 633}, 377, for a recent study of the source).
faints below the visibility level. The green contour denotes the 6-10 keV 
ASCA excess region (see the text above).}
\end{figure}

{\bf Note added on the submission to astro/ph on the 28th Dec 2005.} 
After the submission of the present paper to A\&A on
the 1st of September 2005 two other relevant studies of hard emission 
from NGC 6334 appeared at arXiv.org. These were astro-ph/0510338 
by L.Bassani et al. and astro-ph/0510737 by Y.Ezoe et al. 

\end{document}